\documentclass[%
amsmath,amssymb,
amsfont,
 aps,
 prl,
 longbibliography,
 lengthcheck,%
]{revtex4-1}

\usepackage{amssymb}
\usepackage{amsmath}
\usepackage{graphicx}
\usepackage{epsfig}
\usepackage{mathtools}
\usepackage{dsfont}
\usepackage{natbib}
\def\bea{\begin{eqnarray}}
\def\eea{\end{eqnarray}}

\def\ba{\begin{equation}\begin{array}{c}}
\def\ea{\end{array}\end{equation}}
\def\be{\ba\displaystyle}
\def\ee{\ea}

\begin{document}


\title{The fate of a gray soliton in a quenched Bose-Einstein condensate}

\author{O. Gamayun$^{1,2}$}
\author{Yu. V. Bezvershenko$^{2,3}$}
\author{V. Cheianov$^{1}$}
\affiliation{$^1$ Instituut-Lorentz, Universiteit Leiden, P.O. Box 9506, 2300 RA Leiden, The Netherlands}
\affiliation{$^2$ Bogolyubov Institute for Theoretical Physics, 14-b Metrolohichna str., Kyiv 03680, Ukraine}
\affiliation{$^3$ National University of Kyiv-Mohyla Academy, 04070 Kyiv, Ukraine}

\begin{abstract}

We investigate the destiny of a gray soliton in a repulsive one-dimensional Bose-Einstein condensate undergoing a sudden quench of the non-linearity parameter.
The outcome of the quench is found to depend dramatically on the ratio $\eta$ of the final and initial values of the speed of sound. For integer $\eta$
the soliton splits into exactly $2\eta-1$ solitons. For non-integer $\eta$ the soliton decays into multiple solitons and Bogoliubov modes.
The case of integer $\eta$ is analyzed in detail. The parameters of solitons in the out-state are found explicitly. Our approach exploits
the inverse scattering method and can be easily used for the similar quenches in any classical integrable system.

\end{abstract}
\maketitle


\date{\today}

Ultracold atomic systems have long been attracting the attention of research laboratories
for the investigation of nonlinear phenomena \cite{NONLINEAR}. One of the quintessential objects in nonlinear dynamics
is the soliton, a topologically protected particle-like solitary wave-packet \cite{Rajaraman}.
Theoretical efforts to understand the general properties of solitons such as stable propagation and robustness
in collision processes resulted
in a powerful and elegant mathematical framework known as the inverse scattering theory \cite{Faddeev}. This theory has been successfully applied to many
systems, such as optical fibers, Josephson junctions, molecular systems, and shallow water \cite{optic,*jj,*Dav}.
In ultracold gas systems
solitons were observed experimentally in Bose-Einstein condensates (BECs) \cite{PhysRevLett.83.5198}, their particle nature was confirmed and their collisions were explored \cite{Becker,PhysRevLett.101.130401,Stellmer2008}. One of the great advantages of ultracold gas systems is the high degree of variability of the parameters.
In particular, a lot of experiments have been devoted to the processes resulting in a sudden change of one or several parameters, known as a quench \cite{Greiner,Kinoshita2006,Haller2009,Gring,Trotzky,Ferrari}. These experiments triggered numerous theoretical investigations of the non-equilibrium dynamics of the system after the quench, its relaxation, and the properties of the corresponding steady state (see \cite{Cardy0,Dj,Polkovnikov2011,cauxEssler,Frantzeskakis2010,Konotop2}, and references therein). Quenches from the states possessing nontrivial topological properties remain a largely unexplored area. The importance of topological characteristics of the state for the outcome of the quench was discussed \cite{Dj,Campo1,india}. In particular, it was noticed that the decay of an initial state into a state with a different set of topological quantum  numbers is impossible, a phenomenon dubbed topological blocking in \cite{india}.

In the quasiclassical description of a one-dimensional (1D) BEC a global topological constraint on the motion of the system is imposed by a phase difference between the end points of a trap. In the present work we address the effects of a quench on a state of a repulsive 1D BEC, in which phase difference is saturated by one gray soliton. The quench is a sudden change of the interaction coupling constant, which can be achieved by manipulating either the transverse trapping frequency or the value of scattering length with an external magnetic field \cite{bloch2008many}.
The effect of a quench turns out to be nontrivial already in the quasiclassical (Bogoliubov) limit. In order to maintain the phase shift the soliton
decays into multiple excitations that may be either solitons or Bogoliubov modes. The outcome of the quench is conveniently expressed in terms of the ratio of the final to the initial values of the speed of sound in the condensate, which we denote by $\eta$.
Our main prediction is that when $\eta$ is integer the initial soliton decays into $2\eta-1$ solitons only. The parameters of
these solitons are found explicitly, see Eqs. \eqref{theta2}, \eqref{velocity0}, \eqref{velocity}.
At non-integer $\eta$ the final state is more complex containing a mixture of solitons and Bogoliubov modes.

In the Bogoliubov limit a 1D BEC can be described by the classical field model of the condensate wave function $\Psi(x,t)$ \cite{Pitaevskii2003,Pethick2008}, even though rigorous condensation does not occur \cite{bloch2008many,Lect2004,Castin}, so hereafter we will address to our system as a quasicondensate. The wave function satisfies the Gross-Pitaevski equation which in 1D is also called the nonlinear Schrodinger Equation (NLSE)
\be\label{NLSE1}
i\frac{\partial \Psi}{\partial t} = -\frac{\partial^2\Psi}{\partial x^2} + \frac{c_s^2}{2}\left(|\Psi|^2-1\right)\Psi.
\ee
Here we have put $\hbar=2m=1$ and the function $\Psi(x,t)$ is normalized such that in the ground state $\Psi(x,t) = 1$. In these units the nonlinearity parameter
coincides with the speed of sound $c_s$ in the quasicondensate \cite{Astrakharchik2004}. Eq. \eqref{NLSE1} describes repulsive BEC ($c_s^2>0$) at finite density.
The nonlinearity of the NLSE suggests natural units of length and time: the healing length $\xi = 1/c_s$ and the correlation time $\tau= \xi/c_s=1/c_s^2$.
The natural length scale, $L$, which defines the limits of the applicability of the above classical description, can be determined from the logarithmic behavior of the phase correlation function \cite{Pitaevskii2003,Castin}
\be\label{L}
L = \xi e^{2\pi \xi \rho_0}\,,
\ee
where $\rho_0$ is 1D BEC density.

The quench corresponds to a sudden change
\be\label{quench}
c_s\to \tilde{c}_s= \eta c_s,\,\,\,\, \xi \to \tilde{\xi}_s = \xi/\eta .
\ee
It can be achieved by either varying the coupling constant of the system $g$ or changing its density $\rho_0$, as $c_s \sim \sqrt{g\rho_0}$.
We assume that before and after the quench all scales relevant to our problem are much smaller than $L$ in Eq. \eqref{L}, so we can safely describe our system by Eq. \eqref{NLSE1}.

\begin{figure}[h]
\includegraphics[width=\linewidth]{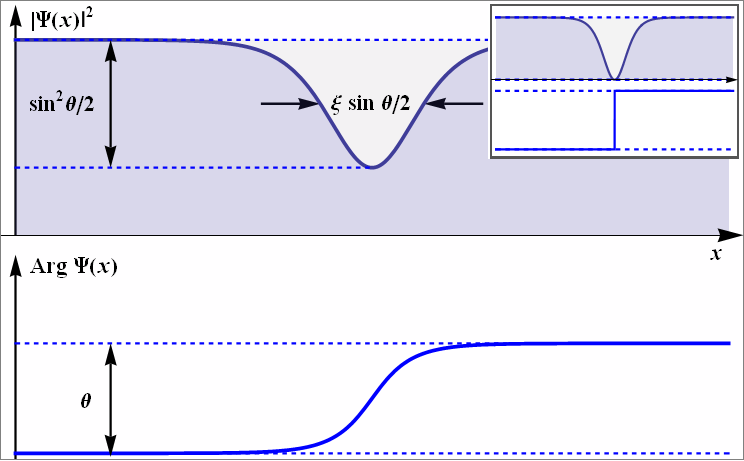}
\caption{\label{Fig1} One-soliton solution.
Upper panel shows the density distribution $|\Psi(x)|^2$, and lower panel shows the phase shift ${\rm Arg}~\Psi(x)$ across the soliton. The special case of  $\theta=\pi$, called the dark soliton, is shown in the inset.}
\end{figure}

The classical ground state remains unaffected by the quench \footnote{in quantum case this is no longer true (see for example \cite{PhysRevA.89.033601})}, therefore we focus our attention on a single soliton, which is the simplest topologically nontrivial excitation. We note that Eq. \label{NLSE} is invariant under a global gauge transformation $\Psi\to \Psi e^{i\theta}$. A soliton is a domain wall separating two vacua with different value of the phase parameter $\theta$.
\be\label{Boundary}
\Psi(x\to-\infty)\to 1,\quad\Psi(x\to +\infty)= e^{i\theta}.
\ee
The generic form of the one-soliton solution is \cite{Faddeev}
\be\label{1sol}
\Psi(x,t) = \frac{1+e^{i\theta}\exp\left(\frac{x-x_0-v(\theta)t}{W(\theta)}\right)}{1+\exp\left(\frac{x-x_0-v(\theta)t}{W(\theta)}\right)},
\ee
where $x_0$ is the initial position of the soliton nodal point and
\be\label{v}
W(\theta) = \frac{\xi}{\sin\theta/2},\,\,\,\,\,v(\theta) = - c_s \cos \frac{\theta}{2}
\ee
are the width and the speed of the soliton, respectively.
Further, without a loss of generality we assume $x_0=0$.
The density and the phase distribution of the soliton at $t=0$ are shown in Fig. \ref{Fig1}.

The density distribution is a single dip of depth $\sin^2\frac{\theta}{2}$ and characteristic width $W(\theta)$.
The soliton expels the amount of matter having the equivalent "volume":
\be\label{dV}
\delta V(\xi,\theta)\equiv\int\limits_{-\infty}^{\infty} dx (1-|\Psi(x,t)|^2) = 2\xi\sin\frac{\theta}{2}.
\ee
We see that the shape and the velocity of a soliton are completely determined by a single parameter $\theta\in [0,2\pi)$, which is the phase difference between the asymptotical regions \eqref{Boundary}.
Smaller values of the phase $\theta$, for $\theta<\pi$, correspond to the faster, shallower and wider soliton, and oppositely for $\theta>\pi$. The maximum speed of a soliton is the speed of sound, corresponding to $\theta=0$.
The special case $\theta=\pi$ is called a dark soliton. It corresponds to the deepest soliton, which does not move at all.
Its phase has a jump through the nodal point of the soliton (see the inset in Fig. \ref{Fig1}).

Recent works show that a quantum soliton state can be defined beyond the Bogoliubov limit by taking into account many-body effects. The obtained dark soliton-like density profile remains stable for a long time \cite{PhysRevLett.108.110401,SatoArxiv,PhysRevLett.112.040402,PhysRevLett.103.140403}. In our semiclassical model solitons are stable for all periods of time, unless the whole system is driven externally. This seems to be valid approximation.

{\it Statement of the problem.} We consider the initial state as a single soliton with the phase shift $\theta$ and assume that the quench \eqref{quench} is performed instantaneously.
It is easily seen that the wave function describing a soliton before the quench no
longer corresponds to a single soliton after the quench.
Indeed, while the phase shift $\theta$ remains unaffected by the quench, the expelled volume of a single soliton solution would have
to instantaneously shrink by a factor $\eta$ as can be seen from Eq.~\eqref{dV}. This is only possible if the excess volume is
carried away by additional excitations resulting from the quench. Our task is to characterize these excitations.

\begin{figure*}[ht]
\includegraphics[width=\linewidth]{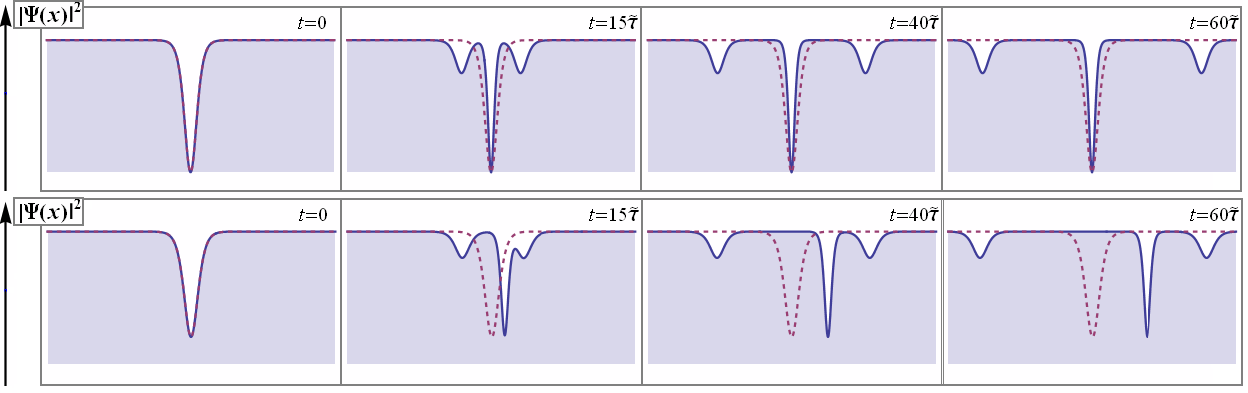}
\caption{\label{Fig2} Dynamics of the one-soliton density profile after the quench $c_s\rightarrow2c_s$. The initial distribution is shown by dashed line.
The evolution after the quench is shown by solid line. Time is measured in units of correlation time $\tilde{\tau}=\tilde{\xi}/\tilde{c}_s$ in the quenched quasicondensate. Top panel shows the splitting of the dark soliton. This splitting results in a symmetric density distribution consisting of one dark soliton and two grey solitons moving with equal speeds in opposite directions.
Bottom panel shows how a gray soliton splits into three solitons with different nonzero velocities. The evolution is calculated analytically from Eqs. \eqref{p1}, \eqref{p2}, \eqref{p3} with the phases given by \eqref{theta2}.}
\end{figure*}

{\it Results.} We find that when $\eta$ is integer, a soliton decays into $(2\eta-1)$-solitons with the phase shifts
\be\label{theta2}
\theta_0 = \theta,\quad
\theta_k^+ = 2\arcsin \frac{k}{\eta}\sin\frac{\theta}{2},\quad
\theta_k^- = 2\pi-\theta_k^+
\ee
where $k=1,2,\dots, \eta -1$. Using Eq. \eqref{v} we see that $\eta$ of these solitons are moving
in the same direction as the initial soliton with the velocities
\be\label{velocity0}
v^+_k = - \tilde{c}_s \sqrt{1-\frac{k^2}{\eta^2}\sin^2\frac{\theta}{2}},\,\,\,\ k=1,2,\dots\eta
\ee
and $\eta-1$ solitons moving in the opposite direction with the velocities
\be\label{velocity}
v^-_k = \tilde{c}_s \sqrt{1-\frac{k^2}{\eta^2}\sin^2\frac{\theta}{2}},\,\,\,\ k=1,2,\dots\eta-1.
\ee
In the case of a dark soliton decay, one soliton remains dark (not moving), while other solitons form $\eta-1$ pairs symmetrically running away from the dark soliton.
We recall that $\tilde{c}_s$ is the speed of sound in the quenched quasicondensate \eqref{quench}.
Experimentally, the solitons can be resolved at times greater the typical separation time $t>t_{\rm sep}$ which can be estimated as
\be\label{time}
t_{\rm sep} = \tilde{\tau}\frac{8\eta^2}{2\eta-1}\frac{\sqrt{1-(1-\eta^{-1})^2\sin^2\theta/2}}{\sin^3\theta/2}.
\ee
Here $\tilde{\tau} = \tilde{\xi}/\tilde{c_s}$ is the correlation time in the quenched quasicondensate.
The exemplary plots for $\eta=2$ soliton decay are shown in Fig. \eqref{Fig2}.

The idea of the generation of solitons by the modulation of the scattering length (nonlinearity constant) was proposed by different groups of authors some time ago.
The idea of spatially modulated scattering length  was considered in \cite{PhysRevLett.95.153903,PhysRevA.74.013619,PhysRevA.74.053614,PhysRevA.78.013611}, while numerical investigation of the {\it continuous} time dependence of switching on spatial variance was considered in \cite{Wang20103863,Konotop3}.

{\it Derivation.}  Before deriving the above results we first demonstrate its consistency by comparing expelled volumes before and after the quench. Indeed,
\be\nonumber
\delta V \left(\frac{\xi}{\eta},\theta\right) + \sum\limits_{k=1}^{\eta-1} \delta V \left(\frac{\xi}{\eta},\theta^+_k\right) +
\sum\limits_{k=1}^{\eta-1} \delta V \left(\frac{\xi}{\eta},\theta^+_k\right) \\
\displaystyle =2\xi\sin\frac{\theta}{2}\left(\frac{1}{\eta}+2\sum\limits_{k=1}^{\eta-1}\frac{k}{\eta^2}\right) = \delta V(\xi,\theta).
\ee
To prove the statements \eqref{theta2}, \eqref{velocity0}, \eqref{velocity} and to generalize to the case of arbitrary $\eta$ we use the inverse scattering transformation (IST) method \cite{Faddeev,ZSH}. This method can also be applied to a range of nonlinear equations such as the Korteweg-de Vries equation,
the Sine-Gordon equation, the Toda lattice equation and others equations all possessing topologically nontrivial solutions (see, for instance, \cite{Faddeev,Babelon} and references therein).
The IST method exploits an auxiliary linear problem, which for Eq. \eqref{NLSE1} is written as
\be\label{linear}
\frac{dF}{dx}  = \frac{1}{2}\left(
 \begin{array}{cc}
        -i\lambda & c_s \bar{\Psi}(x,t) \\
        c_s \Psi(x,t) & i\lambda \\
      \end{array}
    \right)F \equiv U F.
\ee
Here $\Psi(x,t)$ is the value of the field at time $t$ and $\lambda$ is called the spectral parameter.
The idea of the IST method is to consider the linear problem \eqref{linear}
as a scattering problem and to express the field variables $\Psi(x,t), \bar{\Psi}(x,t)$ in terms of the scattering data encoded in the transfer-matrix
\be\label{T}
T(\lambda) = {\rm Pexp}\left(\int\limits_{-\infty}^{\infty}dx U(x,\lambda) \right)\equiv \left(
\begin{array}{cc}
                                                                                              a(\lambda) & \bar{b}(\lambda) \\
                                                                                              b(\lambda) & \bar{a}(\lambda) \\
                                                                                            \end{array}
\right).
\ee
Here $P$ stands for the Dyson path ordering operator.
The scattering data consists of the functions $a(\lambda)$ and $b(\lambda)$ for $\lambda^2 >c_s^2$ and
the discrete set $\lambda_j\in [-c_s,c_s]$ which are the zeroes of the analytically continued $a(\lambda)$ together with associated variables $q_j$.
It turns out that the dynamics of the scattering data that follows from the Eq.~\eqref{NLSE1} is trivial \cite{Faddeev}
\be\label{12}
a(\lambda,t) = a(\lambda,0),\,\,\,\,\,\, b(\lambda,t) = e^{-i\lambda\sqrt{\lambda^2-c_s^2}t}b(\lambda,0),
\ee
\be\label{13}
\lambda_j(t) = \lambda_j(0),\,\,\,\,\,\,\, q_j(t) = q_j(0)+ t\lambda_j\sqrt{c_s^2-\lambda_j^2}.
\ee
The time dependent solution of Eq.~\eqref{NLSE1} is found by the IST from the scattering data \eqref{12}, \eqref{13} to the field variables.
The IST in a generic situation reduces to the solution of a linear integral equation called the Gelfand-Levitan-Marchenko equation (GLM).
A reflectionless transfer matrix, $b(\lambda)=\bar{b}(\lambda)=0$, describes a solution where only solitons are present. In this case the GLM equation
can be solved explicitly.  To describe such a solution we use the parametrization $\lambda = c_s\cosh \varphi$ and
$\lambda_j =  - c_s \cos(\theta_j/2)$, where each phase $\theta_j\in[0,2\pi]$. The general diagonal element of the transfer-matrix \eqref{T} corresponding to the $N$-soliton solution is given by
\be\label{a2}
a(\lambda) = e^{\frac{i\theta}{2}}\prod\limits_{j=1}^N \frac{e^\varphi+ e^{-\frac{i\theta_j}{2}}}{e^\varphi+ e^{\frac{i\theta_j}{2}}}.
\ee
The boundary conditions \eqref{Boundary} are ensured by the $\theta$-condition
\be
\theta = \sum\limits_{j=1}^N \theta_j\,\,\,({\rm mod} \,2\pi).
\ee
The result of the IST for such a transfer matrix can be written
in the following form \cite{Faddeev}
\be\label{p1}
\Psi(x,t) = \frac{{\rm det}(1+\tilde{A})}{{\rm det}(1+ A)}
\ee
where
\be\label{p2}
A_{jk} = \frac{2i\sqrt{m_jm_k}}{e^{i\theta_k/2}-e^{-i\theta_j/2}},\quad \tilde{A}_{jk} = A_{jk}e^{i\frac{\theta_j+\theta_k}{2}}
\ee
and
\be\label{p3}
m_j = \sin\frac{\theta_j}{2}\prod\limits_{k\neq j} \frac{\sin\frac{\theta_j+\theta_k}{4}}{|\sin\frac{\theta_j-\theta_k}{4}|}
e^{
\frac{x}{\xi}\sin\frac{\theta_j}{2}-q_j(t)
}.
\ee
For $N=1$ Eq. \eqref{p1} reduces to the one-soliton solution \eqref{1sol}. In the $t\to\infty$ limit the solution \eqref{p1} consists of
$N$ isolated solitons, each characterized by its own phase shift $\theta_j$.

Next we turn to the analysis of the quench \eqref{quench} of a system containing one soliton. We use the wave function \eqref{1sol} as the initial condition for the NLSE with the quenched speed of sound. To this end, we substitute the wave function \eqref{1sol} into the auxiliary linear problem \eqref{linear} and calculate the transfer-matrix \eqref{T}.
Introducing variable $z= \tanh \frac{\nu (x-x_0)}{2}$ we can present Eq. \eqref{linear} in the form of
\be
\frac{dF}{dz} = \left(\frac{A_-}{1-z}+\frac{A_+}{1+z}\right)F,
\ee
where:
\be
A_- = \frac{\omega}{2\nu}\left(
 \begin{array}{cc}
        \displaystyle -i\cosh\varphi & e^{-i\theta} \\
        \displaystyle e^{i\theta} & \displaystyle i\cosh\varphi \\
      \end{array}
    \right)\\ \displaystyle
A_+ = \frac{\omega}{2\nu}\left(
 \begin{array}{cc}
        \displaystyle -i\cosh\varphi & 1 \\
        \displaystyle 1 & \displaystyle i\cosh\varphi \\
      \end{array}
    \right)
\ee
This equation is of hypergeometric type and the resulting transfer-matrix reads
\be\label{19}
T = \frac{\rho}{\eta}\left(
 \begin{array}{cc}
         \alpha \gamma^- & \displaystyle -i\gamma \\
         -i\gamma & \displaystyle -\bar{\alpha}\gamma^+ \\
      \end{array}
    \right),
\ee
where
\be
\rho = \frac{i\eta\sinh \varphi}{\sin \frac{\theta}{2}},\quad \alpha = \frac{\sin\frac{\theta}{2}}{\sin\left(\frac{\theta}{2}+i\varphi\right)}
\ee
and
\be\label{21}
\gamma = \frac{\Gamma(-\rho)\Gamma(\rho)}{\Gamma(-\eta)\Gamma(\eta)},\quad
\gamma^{\pm} =
\frac{\Gamma(\pm\rho)\Gamma(\pm\rho)}{\Gamma(\pm\rho-\eta)\Gamma(\pm\rho+\eta)}.
\ee
A remarkable consequence of \eqref{21} is that at integer $\eta$ the transfer-matrix is reflectionless, therefore only solitonic excitations are present.
The coefficient $a(\lambda)$ takes the form \eqref{a2} with the set of the phases from Eq.~\eqref{theta2}
and the initial value $q_i(0)=0$ for all $(2\eta-1)$ parameters. After some time the $(2\eta-1)$-solution splits into one-soliton solutions
that moves with velocities \eqref{velocity0} and \eqref{velocity}. The evolution after a particular quench $\eta=2$ is shown at Fig. \ref{Fig2}.
We see that shallower and therefore quicker solitons run away from the initial soliton, which itself grows thinner.
This way, the soliton separation time can be estimated as a time
needed for the nodal point of the soliton to travel further than the width of the adjacent soliton. This gives estimation \eqref{time}.

In the case of noninteger $\eta$ the off-diagonal element of the transfer-matrix $b(\lambda)$ is non-zero, therefore apart from the solitons,
other solutions are present which at small $b(\lambda)$ are just linear waves called the Bogoliubov modes.
Unfortunately, for general $b(\lambda)$ the GLM equation does not possess an analytic solution and we do not expect $\Psi(x,t)$ to have a simple form.

One can consider asymptotic one-soliton solutions as separate particles, then a multisoliton solution describes an effective "interaction" between these particles \cite{zs}. There are plenty of effects which are expected to arise due to this "interaction" \cite{PhysRevLett.110.035303,PhysRevLett.113.036403}.
We expect that some of these effects will reveal themselves in the nontrivial thermodynamic and transport properties of the quasi-condenstate.

In conclusion, we have considered the decay of a single soliton in a Bose-Einstein condensate after the instantaneous change of the nonlinearity parameter. We have found
the conditions under which the soliton splits into an integer number of solitons.
Our findings are based on the inverse scattering transformation method, therefore, this result can be easily extended to
physical systems described by other integrable equations such as the Sine-Gordon and the KdV equations.
Indeed, all those equations are described by the $2\times2$ Lax matrix (in our case it is $U$ in Eq. \eqref{linear}) and possess soliton type solutions. This means that direct scattering problem \eqref{linear} will always have a solution in terms of hypergeometric functions for any values of parameters (for such linear systems, see for instance \cite{Jimbo1982} and recent related discussion in \cite{painleve1,painleve2}).

Although we have considered a homogeneous quasicondensate, we expect that our result will remain valid in the presence of the axial trap potential as long as the width of the soliton is much smaller than the length of the cloud. Indeed, this can be considered in terms of a local density approximation method (e.g. \cite{Astrakharchik2013}) or in terms of a quasiclassical approach \cite{PhysRevA.68.043614}.
We note that the protocol proposed in this work can be used for the creation of multiple solitons in a one-dimensional quasicondensate for the investigation of their joint dynamics. It is also interesting to consider the use of a similar protocol in fiber optic devices for the replication of optical signals.

The initial density profile of a soliton in a pre-quenched quasicondensate can be considered as a special type of initial conditions. From this point of view the analogous problem in the case of bright solitons was considered in \cite{Miles}.

Recently, we became aware of another work \cite{Gromov2}, where the same problem was addressed using a different technique.

\acknowledgements{The authors are grateful to Andrey Gromov for fruitful discussions, for bringing to our attention the work \cite{Gromov2}
and his wonderful talk at APS March Meeting that inspired this investigation \cite{Gromov}. We thanks Benjamin Doyon, Vincent Caudrelier and
Fabio Franchini for bringing our attention to the Ref. \cite{Miles}.
The present work was supported by the ERC grant 279738-NEDFOQ, and partially by the Program of fundamental research of the physics and astronomy division of NASU, and by project 01-01-14 of NASU.}

\bibliographystyle{apsrev4-1}
\bibliography{Quench}

\end{document}